\documentclass[aps,prl,superscriptaddress,showkeys,showpacs,twocolumn,longbibliography,nofootinbib]{revtex4-1}

\usepackage[colorlinks, linkcolor=red, anchorcolor=green, citecolor=blue]{hyperref}
\usepackage{amsmath}
\usepackage{amssymb}
\usepackage{graphicx}
\usepackage{float}
\usepackage{color}
\usepackage{braket}
\usepackage{orcidlink}
\usepackage{comment}
\usepackage{diagbox}
\graphicspath{{figures}}
\allowdisplaybreaks[4]

\newcommand{\bp}{\boldsymbol{p}}
\newcommand{\bk}{\boldsymbol{k}}
\newcommand{\basis}{\mathcal{P}}

\begin{document}

\title{Thermal Spectra Without Detailed Balance}
\author{Xingjian Lu}
\email[]{lus21@mails.tsinghua.edu.cn}
\affiliation{Department of Physics, Tsinghua University, Beijing 100084, China}

\author{Shuzhe Shi}
\affiliation{Department of Physics, Tsinghua University, Beijing 100084, China}
\begin{abstract}
A thermal spectrum is often taken as a signature that the emitted probe has reached detailed balance with the surrounding medium. We show that this interpretation is not generally valid by studying how the microscopic emission kernel determines the macroscopic spectrum. In $3+1$ dimensions, a simple thermal spectrum can be generated without probe thermalization when the relevant kernel belongs to a thermally degenerate class. A representative case is realized when the differential cross section depends on the scattering angle but carries no additional dependence on the Mandelstam variable $s$, as in low-energy Thomson scattering. Our results provide a kernel-based criterion for distinguishing genuine probe--medium exchange equilibrium from thermal spectra produced by the structure of the emission kernel itself.

\end{abstract}

\maketitle

\emph{Introduction.}--- 
Thermal spectra are often used as thermometers~\cite{Planck:1901tja,Rybicki:2004hfl,Shen:2013vja,Rapp:2014hha,ALICE:2015xmh,Shen:2016odt}. 
In blackbody radiation, the Planck spectrum of photons reflects thermal equilibrium between matter and the radiation field~\cite{Planck:1901tja,Rybicki:2004hfl}. 
The temperature of a blackbody, such as the surface of a star, can therefore be extracted by fitting the photon spectrum with an equilibrium distribution~\cite{Mather:1993ij,Fixsen:1996nj, Rybicki:2004hfl,Fixsen:2009ug}.
More generally, if a particle species has reached detailed balance with a medium before decoupling, its momentum distribution is expected to reflect the temperature of the system.

For direct photons in high-energy nuclear collisions, the situation is different. 
After being produced, photons interact only weakly with the surrounding medium and tend to escape without substantial final-state rescattering. 
Their spectrum therefore does not have to be the equilibrium momentum distribution of a thermalized photon gas. 
Instead, even when the medium itself is thermal, the photon production rate is controlled by both the thermal distribution of the medium particles and the microscopic photon-production kernel~\cite{Peitzmann:2001mz,Arnold:2001ms,Stankus:2005eq,Shen:2013vja,David:2019wpt}.

This leads to a basic question. 
If a probe is produced in a microscopic process and then escapes from a thermal medium without sufficient momentum exchange, should its spectrum generically be nonthermal? 
Similar questions arise for other penetrating probes, such as dileptons in high-energy nuclear collisions~\cite{McLerran:1984ay,Rapp:2013nxa} and neutrinos emitted from supernovae~\cite{Keil:2002in,Tamborra:2012ac} or nuclear reactors~\cite{Huber:2011wv,Hayes:2016qnu}. 
In this work, we will show that in some processes, the emitted probes can be produced following the thermal distribution, even if they leave the medium without further equilibration. 
We assume that the emitted probes are produced by thermal medium particles and systematically study how the momentum spectra of these probes depend on the production process.
Several related developments in kinetic theory provide useful context~\cite{Blaizot:2017lht,Dash:2020zqx,deBrito:2023tgb,Aniceto:2024pyc,deBrito:2024qow,deBrito:2024vhm,Hu:2024utr,Gangadharan:2025qbp,Wang:2025wyh}. 

To make this statement precise, we classify the production kernels $\mathcal{K}$, or equivalently the differential-cross-section structures entering them, according to the spectral form they generate in an equilibrium medium.

The first class consists of \textit{thermally degenerate kernels}. 
For these kernels, the production rate can already take the simple thermal form
\begin{align}
R_c[\mathcal{K}](k)
=
C\,k^2 f_{\mathrm{eq}}(E_k/T)
\end{align}
in an equilibrium medium, even without probe--medium exchange equilibration. 
Here $R_c[\mathcal{K}](k)$ denotes the momentum-differential production rate of the emitted probe in the local rest frame of the medium. 
The coefficient $C$ is independent of $k$, $E_k$ is the probe energy, and $T$ is the temperature of the medium. 
The factor $k^2$ is the Jacobian from the three-dimensional momentum phase space and is therefore not regarded as part of the nontrivial spectral structure. 
The function $f_{\mathrm{eq}}(E_k/T)$ denotes the appropriate equilibrium distribution,
with $f_{\mathrm{eq}}=e^{-E_k/T}$ for classical statistics,
$f_{\mathrm{eq}}=1/(e^{E_k/T}-1)$ for bosons, and
$f_{\mathrm{eq}}=1/(e^{E_k/T}+1)$ for fermions.

For thermally degenerate kernels, observing a simple thermal spectrum is not sufficient evidence that the emitted probe has thermalized with the medium or that detailed balance has been established. 
Instead, these kernels may serve as cleaner diagnostics of the medium itself. 
If they naturally produce a simple thermal spectrum in an equilibrium medium, then deviations from this form can signal medium nonequilibrium with less contamination from intrinsic kernel-induced spectral structure.

The second class consists of \textit{exchange-diagnostic kernels}. 
For these kernels, if the emitted probe has not reached exchange equilibrium with the medium, the production rate generally contains an additional nonthermal structure,
\begin{align}
R_c[\mathcal{K}](k)
=
\Phi(k)\,k^2 f_{\mathrm{eq}}(E_k/T),
\end{align}
where $\Phi(k)$ is a nonconstant function generated by the microscopic production kernel. 
The nonconstant factor $\Phi(k)$ makes the spectrum nonthermal unless probe--medium exchange reaches equilibrium, or at least approximately satisfies detailed balance. 
Thus, exchange-diagnostic kernels can serve as reliable indicators of probe--medium exchange equilibration.

\vspace{3mm}
\emph{Method.}--- 
We consider the production rate of a probe species $c$ associated with a generic kernel $\mathcal{K}$ in an equilibrium medium composed of species $a$ and $b$. For the scattering process
$
a(\bp)+b(\bp')\to c(\bk)+d(\bk'),
$
where $\bp$ and $\bp'$ denote the momenta of the medium particles, $\bk$ denotes the momentum of the observed probe, and $\bk'$ denotes the momentum of the accompanying final-state particle, the total production rate $R_c$ is obtained by integrating the momentum-differential production rate $\mathcal{R}_c(\bk)$ over the probe momentum,
\begin{align}\label{eq:total_production}
R_c[\mathcal{K}]
=
\int_{\bk} \mathcal{R}_c(\bk)[\mathcal{K}] \, .
\end{align}
Here we use the shorthand
$
\int_{\bp} \equiv \frac{d^3\bp}{2p^0 (2\pi)^3} \, .
$
Within the kinetic approach, the production for the produced probe with momentum $\bk$ can be written as
\begin{align}\label{eq:gain_term_production}
\begin{split}
&
\mathcal{R}_c(\bk)[\mathcal{K}]
\\=\;&
\int_{\bp,\bp',\bk'}
\mathcal{K}\,
f_a(E_{\bp}/T)\,f_b(E_{\bp'}/T)\,
\bigl(1 +s_d f_d(E_{\bk'}/T)\bigr) \, .
\end{split}
\end{align}
The kernel $\mathcal{K}$ contains the energy--momentum conserving delta function and the squared transition amplitude. More explicitly, for a given macroscopic scattering channel
$
a(\bp)+b(\bp')\to c(\bk)+d(\bk'),
$
we write
\begin{align}
\mathcal{K}
=
(2\pi)^4
\delta^{(4)}(p^\mu+p'^\mu-k^\mu-k'^\mu)
\,
\sum_{\text{int.}}
\left|
\sum_{\lambda}
\mathcal{M}_{\lambda}
\right|^2 .
\end{align}
Here $\lambda$ labels the different diagrams to the same scattering process. These contributions are summed at the amplitude level before taking the modulus squared, so that their interference terms are retained.  The outer summation $\sum_{\text{int.}}$ denotes the relevant sums and/or averages over internal quantum numbers, such as spin, polarization, color, flavor, or other discrete indices. Possible degeneracy factors, statistical symmetry factors also be absorbed into the definition of the kernel.
The functions $f_a$, $f_b$, and $f_d$ denote the equilibrium distribution functions of the corresponding particle species. 
The factors $\bigl(1+s_i f_i\bigr)$ account for Bose enhancement or Pauli blocking for the final-state particles, where $s_i=+1$ for bosonic species and $s_i=-1$ for fermionic species.

To characterize the momentum dependence of the production spectrum, we expand the momentum-differential production rate of the probe species $c$ in a complete orthonormal basis $\basis_i(\bk/T)$,
\begin{align}
\mathcal{R}_c(\bk/T)[\mathcal{K}]
=
\sum_i f_c(E_{\bk}/T)\,
\basis_i(\bk/T)\,
\mathcal{R}_{c,i}[\mathcal{K}] .
\end{align}
where $f_c$ is the equilibrium distribution function of species $c$. $T$ denotes the temperature of the fluid cell in the local rest frame.
The basis functions are chosen to be orthogonal with respect to the equilibrium weight $f_c$,
\begin{align}
\int d^3\bk\,
f_c(E_{\bk}/T)\,
\basis_i(\bk/T)\,
\basis_j(\bk/T)
=
\mathcal{N}_i\,\delta_{ij},
\end{align}
This representation is useful because a thermal-form spectrum and its non-thermal distortions can be distinguished directly through the pattern of excited spectral modes.

The coefficient of each basis mode in the momentum-differential production rate is obtained by projecting the production rate onto the corresponding basis function,
\begin{align}
\begin{split}
&
    \mathcal{R}^{\,\mathrm{gain}}_{c,i}[\mathcal{K}]
    =\;
    \frac{1}{\mathcal{N}_i}
    \int_{\bp,\bp',\bk,\bk'}
    2 k^0 (2\pi)^3
    \basis_i(\bk/T)\,
    \mathcal{K}\,
\\&
    \quad\times
    f_a(E_{\bp}/T)\,f_b(E_{\bp'}/T)\,
    \bigl(1 +s_d f_d(E_{\bk'}/T)\bigr) \, .
\end{split}
\end{align}

To make the kernel-induced spectral structure more transparent, we work in the classical limit in this paper. 
More explicitly, we neglect Bose enhancement and Pauli blocking factors in the final state and approximate the medium distribution functions by Maxwell--Boltzmann distributions. 
This simplification is not essential for the kernel--spectrum correspondence, but it allows the role of the microscopic kernel $\mathcal{K}$ to be isolated in a particularly simple form.

In the local rest frame of an equilibrium fluid cell, the medium is isotropic. 
Since the observable considered here does not introduce any additional preferred direction, rotational invariance implies that the production spectrum depends only on the magnitude of the probe momentum,
$k \equiv \bk$. 
The momentum dependence is therefore reduced to a radial one.

With the classical Boltzmann weight and the radial momentum dependence, a convenient Laguerre basis is
\begin{align}
    \basis_i(k/T) = \frac{1}{2\sqrt{\pi}} L_i^{(2)}(k/T) ,
    \qquad
    \mathcal{N}_i=\frac{(i+2)!}{i!}.
\end{align}
This basis provides a useful diagnostic of whether the produced spectrum has a thermal form. 
After the production rate is computed, the resulting photon spectrum may have a thermal form whose temperature either coincides with the medium temperature, $T_{\rm eff} = T$, or differs from it, $T_{\rm eff} \neq T$. 
In both cases, a thermal-form spectrum leaves a simple and identifiable pattern in the spectral coefficients.

If the produced photon spectrum is thermal with the same temperature as the medium, $T_{\rm eff}=T$, then, only the zeroth mode is excited:
\begin{align}\label{eq:structure1}
\mathcal{R}_{c,i}[\mathcal{K}]
=
\delta_{i0}\,
\mathcal{R}_{c,0}[\mathcal{K}] \, .
\end{align}
All higher modes with $i\geq 1$ therefore vanish.

If instead the produced spectrum still has a thermal form, but with a different effective spectral temperature $T_{\rm eff}$, then it is not represented by a single mode in the basis defined with medium temperature $T$. 
Using the generating function of generalized Laguerre polynomials\footnote{
This follows from the generating function
\[
\sum_{i=0}^{\infty} z^i L_i^{(2)}(x)
=
(1-z)^{-3}
\exp\left(-\frac{xz}{1-z}\right),
\]
with $x=k/T$ and $z=1-T_{\text{eff}}/T$.
}, this thermal spectrum is expanded as a fixed geometric sequence of modes,
\begin{align}\label{eq:structure2}
\mathcal{R}_{c,i}[\mathcal{K}]
=
2\sqrt{\pi}
\left(\frac{T_{\rm eff}}{T}\right)^3
\left(1-\frac{T_{\rm eff}}{T}\right)^i .
\end{align}
Equations~\eqref{eq:structure1} and \eqref{eq:structure2} provide a convenient criterion for identifying whether the produced spectrum has a thermal form.

For a fixed $2\to2$ process with on-shell external particles, the kernel depends on two independent Mandelstam variables. In the center-of-mass frame, it is therefore convenient and complete to parametrize the kernel as
$
\mathcal{K}=\mathcal{K}(\sqrt{s},\cos\Theta),
$
where $\Theta$ is the scattering angle between the incoming particle $a$ and the outgoing particle $c$ in the center-of-mass frame.
The angular dependence on $\cos\Theta$ can then be expanded in Legendre polynomials,
\begin{align}
\mathcal{K}(\sqrt{s},\cos\Theta)
=
\sum_{\ell=0}^{\infty}
\mathcal{K}^{(\ell)}(\sqrt{s})\,P_{\ell}(\cos\Theta),
\end{align}
where $P_{\ell}(x)$ denotes the Legendre polynomial of degree $\ell$, and $\mathcal{K}^{(\ell)}(\sqrt{s})$ is the corresponding coefficient.

In the isotropic equilibrium medium considered here, the production rate is an inclusive one-particle spectrum. Therefore the angular dependence of the kernel is projected onto its angular average. As a result, all nonzero angular modes
give no contribution,
$
\mathcal{R}_{c,i}
\!\left[
\mathcal{K}^{(\ell)}(\sqrt{s})P_{\ell}(\cos\Theta)
\right]
=0,
\qquad
\ell\neq 0 .
$
Only the angularly averaged component $\mathcal{K}^{(0)}(\sqrt{s})$ is
relevant for the present calculation.

To make the dependence of the production rate on the kernel explicit, we expand the $\sqrt{s}$-dependence of $\mathcal{K}^{(0)}(\sqrt{s})$ in integer powers of $\sqrt{s}$,
$
\mathcal{K}^{(0)}(\sqrt{s})
\simeq
\sum_{h} c_{0,h}\,(\sqrt{s})^{h}.
$
For each elementary kernel
\begin{align}
\mathcal{K}_{h}(\sqrt{s})=(\sqrt{s})^{h},
\end{align}
we define the corresponding production coefficient
$\mathcal{R}_{c,i}[\mathcal{K}_{h}]$. Since the production coefficient is
linear in the kernel, the result for a general angularly averaged kernel is
obtained by linear superposition once the elementary coefficients are known.

Following the method in Appendices E and F of Ref.~\cite{Lu:2025yry}, the production coefficient for the elementary kernel $\mathcal{K}_{h}$ can be obtained in closed form as
\begin{align}
\begin{split}
&
    \mathcal{R}_{c,i}[\mathcal{K}_{h}]
\\=\;&
    \frac{i!}{(i+2)!}
    \frac{1}{2^{8}}
    \frac{1}{\pi^\frac{7}{2}}
    T^{1+h}
    \Big(\frac{\pi}{2}\Big)^{\frac{1-(-1)^{h}}{2}}
\\&
    \sum_{m=0}^{i}
    \sum_{t=0}^{m}
    (1 + (-1)^{t})(-1)^{m}
    \binom{i+2}{i-m} \frac{1}{m!}
    \binom{m}{t} 
\\&
    \times
    2^{-m} (3  + m+h)! \frac{h!!(t-1)!!}{(3 +h + t)!!}
\end{split}
\end{align}
This formula holds for integer $h>-2$.

This structure becomes a geometric sequence in $i$ only when $h=2$. 
However, in this case the common ratio is zero, so only the first term is nonzero and all higher terms vanish. 
This is precisely a special case of Eq.~\eqref{eq:structure1}. 
Therefore, except for the special cases in which the full rate satisfies the thermal structure in Eq.~\eqref{eq:structure1}, the produced spectrum is not a Boltzmann distribution.

\vspace{3mm}
\emph{Result.}---
Using the kernel decomposition introduced in the previous section, we evaluate the production rate for elementary kernels labeled by the energy power $h$. 
The purpose is to identify which kernel structures can generate a Boltzmann-form production spectrum.

Tables~\ref{tab:eveni-1} and \ref{tab:oddi-1} list representative normalized coefficients for $i=0$--$9$ and provide a compact view of how the spectral coefficients depend on the power $h$. 
The displayed pattern can be summarized by a simple support structure. 
For odd $h$, all coefficients $\mathcal{R}_{c,i}$ are nonzero. 
For positive even $h$, the coefficients have finite support,
\begin{align}
\mathcal{R}_{c,i}\neq 0
\quad \text{only for} \quad
i<\frac{h}{2}.
\end{align}
Thus only a finite number of spectral modes contribute in positive even cases. 
The case $h=0$ should be treated separately: although it is even, it does not follow this finite-support pattern, and all coefficients remain nonzero.

\begin{table}[htbp]
\centering
\scriptsize
\caption{Normalized coefficients for even $h$ with $i=0\text{--}9$.}
\label{tab:eveni-1}
\begin{tabular}{c|cccccccccc}
\hline
\diagbox{$h$}{$i$}
& $0$ & $1$ & $2$ & $3$ & $4$ & $5$ & $6$ & $7$ & $8$ & $9$ \\ \hline
$0$ & $1$ & $\frac{1}{3}$ & $\frac{1}{6}$ & $\frac{1}{10}$ & $\frac{1}{15}$ & $\frac{1}{21}$ & $\frac{1}{28}$ & $\frac{1}{36}$ & $\frac{1}{45}$ & $\frac{1}{55}$ \\
$2$ & $1$ & $0$ & $0$ & $0$ & $0$ & $0$ & $0$ & $0$ & $0$ & $0$ \\
$4$ & $1$ & $-\frac{1}{3}$ & $0$ & $0$ & $0$ & $0$ & $0$ & $0$ & $0$ & $0$ \\
$6$ & $1$ & $-\frac{2}{3}$ & $\frac{1}{6}$ & $0$ & $0$ & $0$ & $0$ & $0$ & $0$ & $0$ \\
$8$ & $1$ & $-1$ & $\frac{1}{2}$ & $-\frac{1}{10}$ & $0$ & $0$ & $0$ & $0$ & $0$ & $0$ \\
$10$ & $1$ & $-\frac{4}{3}$ & $1$ & $-\frac{2}{5}$ & $\frac{1}{15}$ & $0$ & $0$ & $0$ & $0$ & $0$ \\
\hline
\end{tabular}
\end{table}

\begin{table}[htbp]
\centering
\scriptsize
\caption{Normalized coefficients for odd $h$ with $i=0\text{--}9$.}
\label{tab:oddi-1}
\begin{tabular}{c|cccccccccc}
\hline
\diagbox{$h$}{$i$}
& $0$ & $1$ & $2$ & $3$ & $4$ & $5$ & $6$ & $7$ & $8$ & $9$ \\ \hline
$-1$ & $1$ & $\frac{1}{2}$ & $\frac{5}{16}$ & $\frac{7}{32}$ & $\frac{21}{128}$ & $\frac{33}{256}$ & $\frac{429}{4096}$ & $\frac{715}{8192}$ & $\frac{2431}{32768}$ & $\frac{4199}{65536}$ \\
$1$ & $1$ & $\frac{1}{6}$ & $\frac{1}{16}$ & $\frac{1}{32}$ & $\frac{7}{384}$ & $\frac{3}{256}$ & $\frac{33}{4096}$ & $\frac{143}{24576}$ & $\frac{143}{32768}$ & $\frac{221}{65536}$ \\
$3$ & $1$ & $-\frac{1}{6}$ & $-\frac{1}{48}$ & $-\frac{1}{160}$ & $-\frac{1}{384}$ & $-\frac{1}{768}$ & $-\frac{3}{4096}$ & $-\frac{11}{24576}$ & $-\frac{143}{491520}$ & $-\frac{13}{65536}$ \\
$5$ & $1$ & $-\frac{1}{2}$ & $\frac{1}{16}$ & $\frac{1}{160}$ & $\frac{1}{640}$ & $\frac{1}{1792}$ & $\frac{1}{4096}$ & $\frac{1}{8192}$ & $\frac{11}{163840}$ & $\frac{13}{327680}$ \\
$7$ & $1$ & $-\frac{5}{6}$ & $\frac{5}{16}$ & $-\frac{1}{32}$ & $-\frac{1}{384}$ & $-\frac{1}{1792}$ & $-\frac{5}{28672}$ & $-\frac{5}{73728}$ & $-\frac{1}{32768}$ & $-\frac{1}{65536}$ \\
$9$ & $1$ & $-\frac{7}{6}$ & $\frac{35}{48}$ & $-\frac{7}{32}$ & $\frac{7}{384}$ & $\frac{1}{768}$ & $\frac{1}{4096}$ & $\frac{5}{73728}$ & $\frac{7}{294912}$ & $\frac{7}{720896}$ \\
\hline
\end{tabular}
\end{table}

This support structure immediately identifies the Boltzmann-form case. 
For $h=2$, the finite-support condition gives
\begin{align}
i<1,
\end{align}
so that only the zeroth spectral mode contributes. 
Since the zeroth basis mode is $\frac{1}{\sqrt{4\pi}}$, the emitted spectrum has the Boltzmann form $\mathcal{R}_{c}\left[\mathcal{K}_2\right]=\frac{1}{\sqrt{4\pi}}e^{-k/T}\left(\frac{k}{T}\right)^2$.
Since the production rate is not sensitive to the angular dependence of the kernel, we conclude that, for this class of kernels, the production spectrum is given by the Boltzmann distribution as long as the kernel is proportional to $s$ after angular averaging.

This result is nontrivial because it does not rely on subsequent re-scattering or probe-medium thermalization. 
It follows only from the energy dependence of the angularly averaged emission kernel.

\begin{figure}
    \centering
    \includegraphics[width=1.0\linewidth]{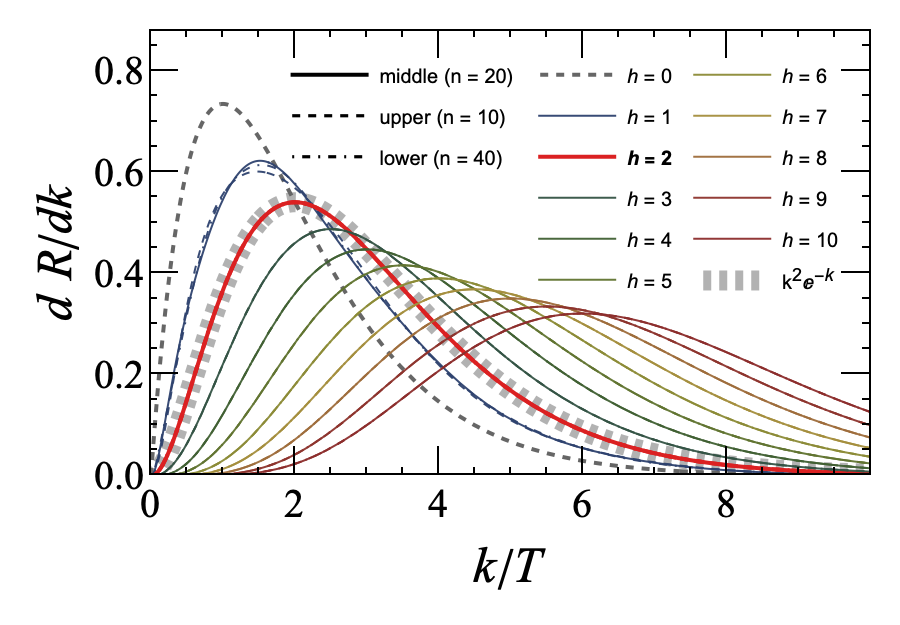}
    \caption{Dependence of the production spectrum on $h$ for $\ell=0$. 
    The spectra are reconstructed by summing the spectral expansion up to order $n$, 
    i.e., including coefficients with $0\leq i\leq n$. 
    The solid, dashed, and dash-dotted lines correspond to $n=20$, $n=10$, and $n=40$, respectively.}
    \label{fig:h_dependent}
\end{figure}

Figure~\ref{fig:h_dependent} shows the normalized production spectra ($\int dk \mathcal{R}_c(k)=1$) in the isotropic sector. 
The figure complements the coefficient analysis by showing how the spectral shape changes away from the Boltzmann-form case. 
As $h$ increases, the peak of the spectrum shifts to larger $k/T$, and the high-momentum tail becomes broader. 
Therefore, larger positive powers of the kernel generate harder emitted spectra.

The $h=2$ spectrum, shown by the thick red curve, overlaps with the gray reference curve for the thermal form $(k/T)^2 e^{-k/T}$. 
This agreement indicates that, for $h=2$, the Boltzmann-type spectrum arises directly from the emission kernel itself, rather than from subsequent probe-medium interactions.
Away from $h=2$, the spectra change systematically: $h<2$ gives softer spectra shifted to smaller $k/T$, while $h>2$ gives harder spectra with broader high-momentum tails. 
Thus the energy dependence of the microscopic kernel controls both the Boltzmann-form case and its deviations.

This special kernel structure, i.e., the process always produces a thermal emitted probe as noted above, is not only formal, but also has a direct physical realization. 
The Thomson scattering,
$
\gamma+e^- \rightarrow \gamma+e^- .
$
in the low-energy limit $E_\gamma \ll m_e$ is such an example.

The full differential cross-section is described by the full Compton-scattering cross section. 
In the center-of-mass frame, the unpolarized differential cross section is~\cite{Klein:1929lcc,Berestetskii:1982qgu}
\begin{align}
\begin{split}
&
    \frac{d\sigma}{d\Omega_{\rm cm}}
=\;
    \frac{\alpha^2}{2s}
    \Big[
    \frac{s-m_e^2}{m_e^2-u}
    +
    \frac{m_e^2-u}{s-m_e^2}
\\&
    +
    \frac{4m_e^2 t}{(s-m_e^2)(m_e^2-u)}
    +
    \frac{4m_e^4 t^2}{(s-m_e^2)^2(m_e^2-u)^2}
    \Big],
\end{split}
\end{align}
with
\begin{align}
t
=
-\frac{(s-m_e^2)^2}{2s}
\left(1-\cos\theta_{\rm cm}\right),
\quad
u=2m_e^2-s-t.
\end{align}
In the low-energy limit, this expression reduces to the Thomson result~\cite{Jackson:1998nia,Rybicki:2004hfl},
\begin{align}
\frac{d\sigma}{d\Omega_{\rm cm}}
=
\frac{\alpha^2}{2m_e^2}
\left(1+\cos^2\theta\right).
\end{align}
where $\alpha$ is the fine-structure constant, and $m_e$ is the electron mass.
Using the standard relation between the differential cross section and the scattering amplitude,
$
\mathcal{K}
=
s\frac{d\sigma}{d\Omega},
$
one finds that the energy-independent Thomson cross section implies
$
\mathcal{K} \propto \frac{s}{m_e^2}\left(1+\cos^2\theta\right).
$
This is precisely the $h=2$ scaling in our kernel decomposition.
Moreover, the angular structure can be expanded as
$
1+\cos^2\theta
=
\frac{4}{3}P_0(\cos\theta)+\frac{2}{3}P_2(\cos\theta).
$
As noted above, only the $\ell=0$ component contributes to the production rate, while the $\ell\neq0$ components do not. Therefore, Thomson scattering provides a direct physical realization of the $h=2$ case that leads to the Boltzmann spectrum discussed above.

\vspace{3mm}
\emph{Summary and Discussion.}---In this Letter, we examined whether a simple thermal emission spectrum of a probe can be regarded as a unique signature of probe--medium exchange equilibration in an equilibrium medium. 
By studying how the emitted spectrum is determined by the microscopic emission kernel, we showed that this interpretation is not generally valid. 
This led us to distinguish between \textit{exchange-diagnostic kernels}, for which the spectrum genuinely reflects probe--medium exchange equilibration, and \textit{thermally degenerate kernels}, for which a simple thermal spectrum can arise even without exchange equilibrium.

A general consequence of our analysis is that, in an equilibrium medium, the emitted probe spectrum is isotropic in the local rest frame. 
Moreover, its spectral shape is controlled only by the angularly averaged part of the differential cross section. 
The remaining energy dependence of the kernel, especially its dependence on the Mandelstam variable $s$, determines the hardness of the spectrum: larger positive powers of $s$ generally lead to harder spectra.

In 3+1 dimensions, the thermally degenerate Boltzmann-form case occurs when the angularly averaged kernel has the relevant $h=2$ structure, or equivalently is proportional to $s$. 
Low-energy Thomson scattering provides a concrete realization of this mechanism.

The main message is therefore that a thermal-like emission spectrum is not, in general, a unique diagnostic of probe--medium exchange equilibration. 
Instead, its interpretation depends on the structure of the emission kernel. 
Our analysis clarifies when a thermal spectral form reflects exchange equilibration and when it follows from kernel-level thermal degeneracy. 
Future extensions to non-equilibrium media will allow one to study how medium anisotropies and nontrivial angular structures of the collision kernel modify the emitted spectrum and reveal additional information about the underlying collision dynamics.

\section*{Acknowledgments} 
The authors are deeply grateful to Pengfei Zhuang for continuous support and encouragement throughout this research. The authors also thank Yunpeng Liu and Xinqi Li for earlier insightful discussions, which provided valuable inspiration for the ideas developed in this work.
This work is supported by NSFC by grant No. 12575143 and by Tsinghua University under grant Nos. 04200500123, 531205006, 533305009. We also acknowledge the support by center of high performance computing, Tsinghua University.
\bibliography{biblio.bib}
\end{document}